\documentclass{article}
\usepackage{spconf,amsmath,graphicx}

\usepackage{enumitem}
\setlist{nosep, leftmargin=14pt}

\usepackage{mwe} % to get dummy images

\title{Cell Maps Representation FOR LUNG Adenocarcinoma GROWTH PATTERNS CLASSIFICATION IN WHOLE SLIDE IMAGES}
 \name{Arwa Al-Rubaian$^{\star}$ \hfill Gozde N. Gunesli$^{\star}$ \hfill Wajd A. Althakfi$^{\dagger}$ \hfill Ayesha Azam$^{\ddagger}$ \hfill  Nasir Rajpoot$^{\star}$ \hfill Shan E Ahmed Raza$^{\star}$}

 \address{$^{\star}$ Tissue Image Analytics Centre, Department of Computer Science, University of Warwick, UK \\$^{\dagger}$Histopathology unit, department of Pathology, king Saud University, Riyadh, Kingdom of Saudi Arabia\\$^{\ddagger}$Department of Histopathology. University Hospitals Coventry and Warwickshire NHS Trust, UK}

\begin{document}
%\ninept
%
\maketitle
\begin{abstract}
Lung adenocarcinoma is a morphologically heterogeneous disease, characterized by five primary histologic growth patterns. The quantity of these patterns can be related to tumor behavior and has a significant impact on patient prognosis. In this work, we propose a novel machine learning pipeline capable of classifying tissue tiles into one of the five patterns or as non-tumor, with an Area Under the Receiver Operating Characteristic Curve (AUCROC) score of 0.97. Our model’s strength lies in its comprehensive consideration of cellular spatial patterns, where it first generates cell maps from Hematoxylin and Eosin (H\&E) whole slide images (WSIs), which are then fed into a convolutional neural network classification model. Exploiting these cell maps provides the model with robust generalizability to new data, achieving $\approx 30\%$  higher accuracy on unseen test-sets compared to current state of the art approaches. The insights derived from our model can be used to predict prognosis, enhancing patient outcomes.
\end{abstract}
\section{Introduction}
\label{sec:intro}

\begin{figure*}[t]
  \centering
  \includegraphics[width=\textwidth]{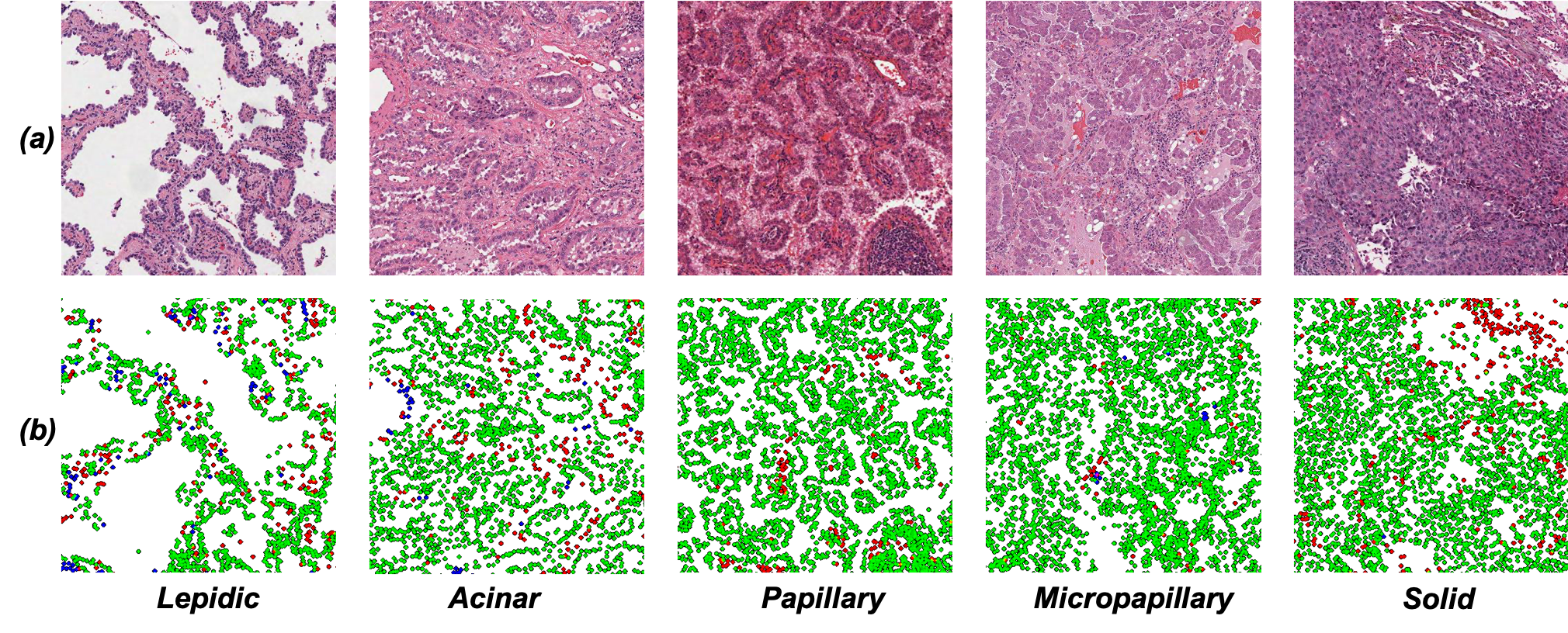}
  \caption{Sample images of lung adenocarcinoma (LUAD)  growth patterns. (a) shows the H\&E image, (b) shows the corresponding cell map, where green represents neoplastic cells, red shows connective cells, and blue shows non-neoplastic cells}
  \label{fig:samplePatters}
\end{figure*}

Lung cancer is one of the most prevalent forms of cancer worldwide, being the second most common cancer (after breast) and the leading cause of cancer-related deaths, accounting for approximately 18\% of all cancer deaths globally \cite{cancerStats}. Around 85\% of all reported lung cancer cases are classified as non-small cell lung cancer (NSCLC), with Lung Adenocarcinoma (LUAD) as its prevalent sub-type. According to the latest World Health Organization (WHO) classification, invasive nonmucinous LUAD grows into five primary histologic growth patterns: lepidic, acinar, papillary, micropapillary, and solid \cite{whoClassification}, examples are shown in Fig.\ref{fig:samplePatters}(a). These patterns can be related to prognosis and effect the patient outcome, with lepidic having the most favorable prognosis, followed by acinar and papillary, whereas solid and micropapillary have been known to have the worst prognosis of all \cite{gradingAdeno}.

As cancer is heterogenous, within the same tumor, LUAD often presents a varied combination of multiple growth patterns. The World Health Organization recommends a three-tiered grading system for lung adenocarcinomas, which involves assessing the predominant histological growth pattern and determining the presence or absence of high-grade patterns (solid and micropapillary). This approach has been shown to provide more accurate information about patient outcomes. However, accurately assessing various growth patterns can be challenging in routine practice due to the presence of mixed patterns, tumor heterogeneity and time constraints. Additionally, quantifying these growth patterns through visual assessment is subjective and can vary between different observers \cite{InterobserverVariability}. Nevertheless, this diagnostic method fails to consider the diverse patterns and their spatial arrangement within the tumor, which could provide more accuracy in diagnosis and improve prognostic predictions for patients. The automation of growth pattern classification using machine learning would be a valuable addition to pathology, enhancing the precision and objectivity of such task while alleviating its labor-intensive and time-consuming nature.

Due to the high dimensionality and large size of Whole Slide Images (WSIs), it is necessary to divide them into smaller tiles before feeding them to deep learning models. As a result, two different approaches for splitting training and testing data have emerged in the literature \cite{Bussola_Marcolini_Maggio_Jurman_Furlanello_2021}: 
\begin{itemize}
\item A \textbf{\textit{WSI-based}} split: Where the splitting of data is done at the WSI level, before the tiling. This ensures that the tiles in the test set originate from WSIs that none of its tiles have been utilized for training. Thus creating a truly unseen test set.
\item A \textbf{\textit{tile-based}} split: Where the WSIs are first divided into tiles, followed by mixing these tiles for subsequent division into training and testing sets. In this approach, training and test sets could have tiles originating from the same WSI. 
\end{itemize}

The classification of LUAD growth patterns is an active area of research. The majority of the machine learning algorithms designed for that task predict a single pattern for each WSI, namely the predominant pattern \cite{pathologist-level_2019}. These methods achieve good accuracies and can be used to assist pathologists in their routine work. However, they do not deliver adequate performance when assessed on the tile level. This might be explained that the aggregation method producing the slide level prediction diminishes the error by neglecting the misclassified tiles. 

Only a limited number of studies \cite{alsubaie_growth_2023,ding_tailoring_2023,sadhwani_comparative_2021, campanella2022hebased}  have developed tile based classifiers. However, in all these studies a tile-based split was adopted. Consequently, a tile in the test set might originate from the same WSI or even be adjacent to a tile used in training the model. This notably enhances the classifier's performance because tiles from the same pattern within a WSI possess substantial visual resemblance, unlike similar pattern tiles from other WSIs. As expected, these models often encounter failure when new data or WSIs that have not been seen before are used for external validation.

In this paper, we introduce a novel machine learning pipeline capable of classifying tissue tiles into one of the five growth patterns or as non-tumor. Our model’s strength lies in its comprehensive consideration of tissue cellular composition, where we first generate cell maps from Hematoxylin and eosin (H\&E) WSIs, which are then fed into a convolutional neural network classification model. Specifically, our contributions are: 
\begin{itemize}
    \item We introduce the concept of cell maps for predicting pathology specific tasks that can add generalizability to current machine learning approaches. Up to our knowledge, this is the first study that uses cell maps representation for tissue classification.
    \item We propose a pipeline for LUAD growth pattern classification that outperforms the current state of the art approaches on unseen test sets, when adopting the WSI-based splitting approach.
\end{itemize}
\begin{figure*}[t]
  \centering
  \includegraphics[width=\textwidth]{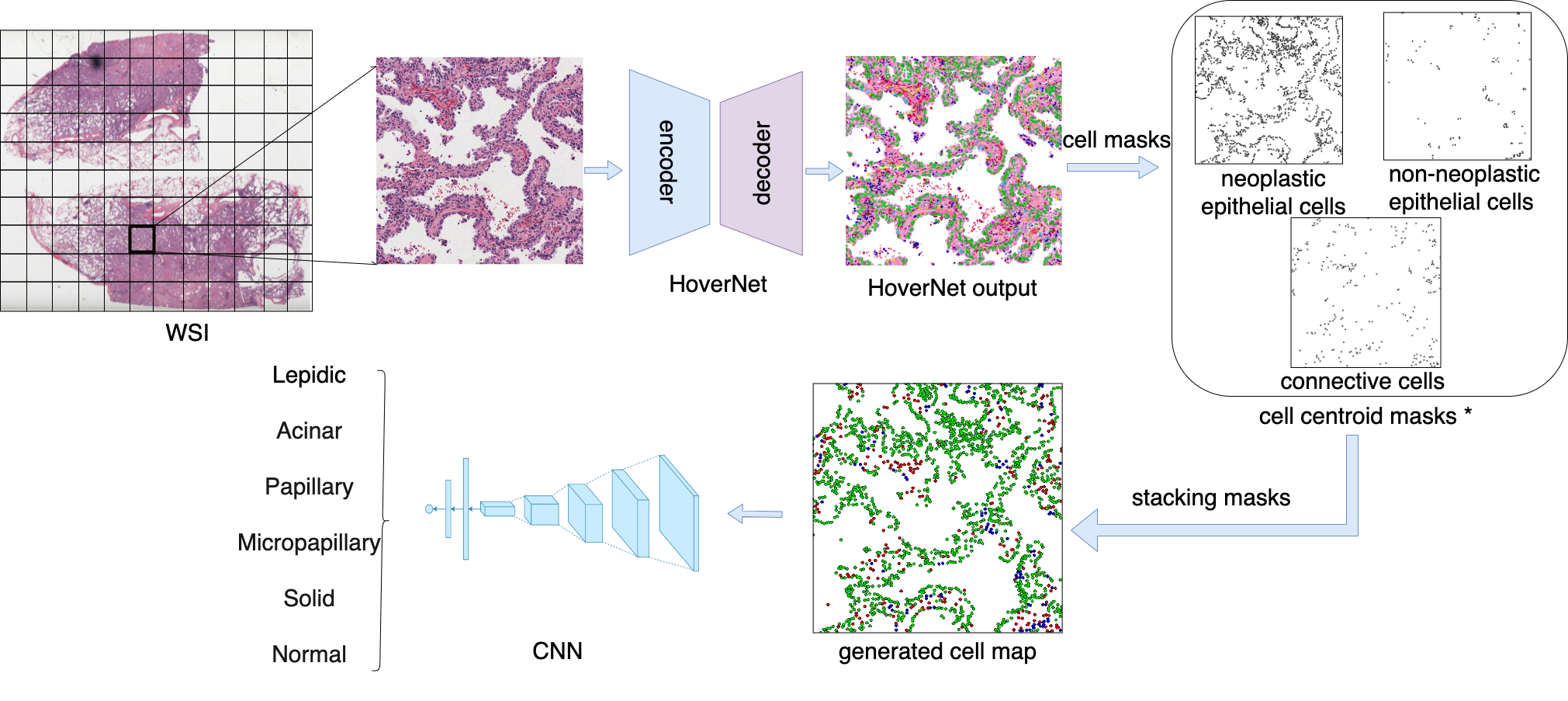}
  \caption{An overview of the proposed Model. First the WSI is divided into tiles, which are then processed using HoverNet to obtain cell locations and types, then a cell map is generated by stacking the masks of three cell types (neoplastic, non-neoplastic, connective). The generated cell map is used as an input to a Convolutional neural network (ResNet50) for growth pattern classification. 
* Colors of the cell masks are inverted for illustration purposes}
  \label{fig:model}
\end{figure*}
\section{The Proposed Method}
\label{sec:format}
The proposed approach can be broken down into two major parts: (1) The construction of the cell maps, (2) training a CNN to predict the different growth patterns. An overview of the proposed method is illustrated in Fig.\ref{fig:model}.
\subsection{Cell maps construction}
\label{ssec:subhead}
We leveraged Hover-Net \cite{graham_hover-net} to find the locations and type of cells in a given tissue. The Hover-Net model was trained on the PanNuke dataset \cite{gamper_pannuke_2019} composed from 19 different tissue types, including lung. We processed the entire WSI at \(20 \times \) magnification. From those resulting nuclei only the neoplastic epithelial, non-neoplastic epithelial, and connective cells were selected. Where the arrangement of the neoplastic cells is the main visual aspect differentiating the different growth patterns. The non-neoplastic cells were used to define the normal tissue, and the connective cells were incorporated to aid in the classification of the papillary pattern; which is defined as tumor papillae arranged around a fibrovascular core. Finally, a binary mask was created for each nuclei class, where a value of 1 was given to a pixel if it corresponds to a nucleus centroid, of that given class, or was in a radius of 4 pixels from it. The three binary masks where then stacked to form a 3-channel cell map image and resized to the dimensions of its WSI at \(5 \times \) magnification. As shown in Fig\ref{fig:samplePatters}(b), the visual distinction between the different patterns is maintained in the generated cell maps.  
\subsection{Growth pattern classification}
\label{ssec:subhead}
We finetuned a ResNet50, pretrained on ImageNet, after adapting the last fully connected layer to predict one of the six classes (lepidic, acinar, papillary, micropapillary, solid, and normal). Cell map tiles of size \(256 \times  256\)  were passed to the network, after applying random horizontal and vertical flipping. Training was optimized using an Adam optimizer and a cross entropy loss function. The model was trained for 20 epochs with a learning rate of 1e-5, which due to the simplicity of the input image representation, we found to be sufficient for the model to converge.
\section{EXPERIMENTS}
\label{sec:pagestyle}
\subsection{Dataset}
\label{ssec:subhead}
The dataset used in this paper consists of 1,034 tiles obtained from 18 WSIs across 12 different centers from the TCGA-LUAD dataset \cite{gdc}. Two pulmonary pathologists independently annotated growth patterns regions on the WSIs. A tile was included if both pathologists opinions agreed on at least 80\% of its area. Each WSI includes on average 2-3 patterns. The most common pattern was solid having 311 tiles, and the least common patterns were papillary and micropapillary having only 54 and 39 tiles, respectively.

\begin{table*}[t]
    \centering
    \begin{tabular}{|c|c|c|c|c|c|c|} \hline  
         &  \multicolumn{3}{|c|}{\textbf {Weak validation}}&  \multicolumn{3}{|c|}{\textbf{Strong validation}}\\ \hline  
         \textbf{Model}&  \textit{AUCROC}&  \textit{F1-macro}&  \textit{accuracy}&  \textit{AUCROC}&  \textit{F1-macro}& \textit{accuracy}\\ \hline  
         \textbf{Proposed Cell maps}&  0.97 ± 0.01&  0.81 ± 0.03&  0.89 ± 0.02 &  \textbf{0.78 ± 0.04}&   \textbf{0.46 ± 0.01}& \textbf{0.63 ± 0.02}\\ \hline  
         H\&E multiscale approach&  1 ± 0.0&  0.96 ± 0.01&  0.97 ± 0.01&  0.58 ± 0.03&  0.28 ± 0.01& 0.29 ± 0.03\\ \hline  
         CoAtNet-1\cite{dai_coatnet_2021}&  0.96 ± 0.02&  0.74 ± 0.05&  0.38 ± 0.02&  0.63 ± 0.04&  0.2 ± 0.04& 0.27 ± 0.03
\\ \hline  
         ResNet50&  1 ± 0.0&  0.96 ± 0.01&  0.48 ± 0.0&  0.63 ± 0.06&  0.25 ± 0.05& 0.32 ± 0.04
\\ \hline  
         AlSubaie et. al\cite{alsubaie_growth_2023} &  0.94 ± 0.01&  0.66 ± 0.03&  0.68 ± 0.04&  0.57 ± 0.04&  0.27 ± 0.02& 0.37 ± 0.08
\\ \hline  
         HoverNet cellular features\cite{graham_hover-net} &  0.79 ± 0.01&  0.39 ± 0.03&  0.59 ± 0.03&  0.55 ± 0.09&  0.23 ± 0.03& 0.34 ± 0.08
\\ \hline 
    \end{tabular}
    \caption{Quantitative comparison of the performance of the proposed method along with other state-of-the-art and baseline methods. The first half uses weak validation approach, when adopting a tile-based data splitting. And the second half uses strong validation by adopting WSI-based splitting  }
    \label{tab:results}
\end{table*}

\subsection{Results}
\label{ssec:subhead}
In this section, we compare our proposed cell maps approach with other state of the art methods and baselines including: 

\begin{enumerate}
    \item \textit{A multiscale approach}: where we proposed to train two identical ResNet-34 models: one on H\&E tiles at \(20 \times \) magnification and the other on H\&E tiles at \(5 \times \) magnification. The final classification result is determined by selecting the prediction with the highest probability among the two models. This approach mirrors the way pathologists typically analyze patterns, where they make an initial decision at lower magnifications and then confirm or alter their decision at higher magnifications.
    \item \textit{CoAtNet-1} \cite{dai_coatnet_2021}: a recently published network architecture combining the self-attention mechanism with convolutions. This architecture achieved the best performance on the GasHisSDB dataset \cite{hu_gashissdb_2022}, a histology dataset of H\&E images for the diagnosis of gastric cancer. We used H\&E tiles at \(20 \times \) magnification as an input to the model.
    \item \textit{ResNet50}: (the architecture used in \cite{ding_tailoring_2023} and \cite{Gertych_Swiderska_2019}) pre-trained on ImageNet fed with H\&E tiles at \(20 \times \) magnification.  
    \item \textit{The approach proposed by AlSubaie et. al.}\cite{alsubaie_growth_2023} :  The input to the model is a 6-channel image, where the tile at \(20 \times \) magnification, and at \(10 \times \) magnification of an H\&E image were stacked and aligned in the center to form the input to a modified ResNet50.
    \item \textit{HoverNet \cite{graham_hover-net} cellular features}: A tile is formulated as a 12-feature vector composed of the count of cells, maximum and minimum distance between cell centroids for each of following cell types : neoplastic, non-neoplastic, connective, and inflammatory cells present in that tile. A support vector machine is then used to classify these feature vectors into the same six classes.
\end{enumerate}

\subsubsection{Weak validation}
\label{sssec:subsubhead}
This validation approach used in current literature involves employing a tile-based data splitting approach, where 5-fold cross validation is applied after all the tiles from different WSIs are mixed. We include the results of this weak validation solely for comparison purposes. We suggest strong validation should be used for testing robustness of the algorithms. All models was trained for 20 epochs with a learning rate of 1e-5.Training was optimized using an Adam optimizer and a cross entropy loss function. Quantitative results are listed in Table\ref{tab:results}.

\subsubsection{Strong validation}
\label{sssec:subsubhead}
For our main experiments, we adopt a WSI-based data splitting approach, where we trained and evaluated each model 5 times. In each trial 6 WSIs were randomly sampled from the dataset to be the unseen test set, ensuring that it includes samples from all the six classes. The remaining slides were divided into tiles and split into 90\% training and 10\% validation set. Training was optimized using an Adam optimizer and a cross entropy loss function. All models was trained for 20 epochs with a learning rate of 1e-5. The average and standard deviation of the performance measures (AUC-ROC, accuracy, and macro F1-score) are reported in Table \ref{tab:results}.

It has been shown in Table \ref{tab:results} that when performing proper validation by adopting a WSI-based data splitting, the proposed cell maps approach performs better on unseen test sets compared to other methods, some of which perform worse than random guessing. Although using the H\&E image directly for classification initially suggests better performance, when properly validating the models via a WSI-based splitting, the performance dramatically drops by over 60\%. Conversely, when employing cell maps, the decline in performance is notably less, dropping by under 20\%. One could argue that cell maps enable CNNs to concentrate on cell morphology without becoming overly fixated on tissue-specific characteristics. 

The proposed cell maps approach was best at distinguishing non-tumor and solid pattern with an average f1-score of 0.98 and 0.97 respectively. On the other hand, the papillary pattern exhibited the highest misclassification error. This could be attributed to the fact that the samples originated from only four WSIs, and given the high heterogeneity of this pattern, the model might not have been exposed to a sufficient number of samples during training. We plan to address this issue by using a larger dataset in the extended version of this work.

\section{CONCLUSIONs}
\label{sec:typestyle}

We proposed a new approach for LUAD growth pattern classification, that outperforms previously proposed approaches in the literature and baseline methods, when evaluated using WSI-based splitting. Additionally, we present a new representation of WSIs, called cell maps, which effectively captures cellular composition in more compact and lighter images. This innovative approach can significantly expedite the training of machine learning algorithms for some pathological tasks. We have also shown that the current validation technique used in the literature to evaluate growth pattern classification models is weak and should be avoided; as it can generate misleading results.
Our future directions include acquiring more annotations to expand our dataset along with testing the model on external cohorts. We aim to project the model predictions on the entire WSI for prognosis and survival analysis.

\section{Acknowledgments}
\label{sec:acknowledgments}

AA is fully funded by the Saudia Arabia Cultural Bureau in London.

\bibliographystyle{IEEEbib}
\bibliography{references}

\end{document}